\documentclass[10pt,conference]{IEEEtran}
\IEEEoverridecommandlockouts
\usepackage{cite}
\usepackage{amsmath,amssymb,amsthm}
\usepackage{algorithmic}
\usepackage{graphicx}
\usepackage{textcomp}
\usepackage{xcolor}
\usepackage[switch]{lineno}
\usepackage{booktabs}
\usepackage{array}
\usepackage{subfigure}
\usepackage{enumitem}
\usepackage{multirow}
\usepackage{mathrsfs}
\usepackage{epstopdf}
\usepackage{caption}
\usepackage{hyperref}
\hypersetup{hidelinks}
\def\BibTeX{{\rm B\kern-.05em{\sc i\kern-.025em b}\kern-.08em
    T\kern-.1667em\lower.7ex\hbox{E}\kern-.125emX}}

\newcommand{\tabincell}[2]{\begin{tabular}{@{}#1@{}}#2\end{tabular}}

\begin{document}

\title{GAHNE: Graph-Aggregated Heterogeneous Network Embedding}

\author{\IEEEauthorblockN{Xiaohe Li, Lijie Wen\IEEEauthorrefmark{1}, Chen Qian, Jianmin Wang}
\IEEEauthorblockA{School of Software, Tsinghua University, Beijing, China\\
Email: \{lixh18,qc16\}@mails.tsinghua.edu.cn, \{wenlj,jimwang\}@tsinghua.edu.cn
}
}
\maketitle
\begin{abstract}
The real-world networks often compose of different types of nodes and edges with rich semantics, widely known as heterogeneous information network (HIN). Heterogeneous network embedding aims to embed nodes into low-dimensional vectors which capture rich intrinsic information of heterogeneous networks. However, existing models either depend on manually designing meta-paths, ignore mutual effects between different semantics, or omit some aspects of information from global networks. To address these limitations, we propose a novel Graph-Aggregated Heterogeneous Network Embedding (GAHNE), which is designed to extract the semantics of HINs as comprehensively as possible to improve the results of downstream tasks based on graph convolutional neural networks. In GAHNE model, we develop several mechanisms that can aggregate semantic representations from different single-type sub-networks as well as fuse the global information into final embeddings. Extensive experiments on three real-world HIN datasets show that our proposed model consistently outperforms the existing state-of-the-art methods.
\end{abstract}

\begin{IEEEkeywords}
Heterogeneous Network, Graph Convolution, Graph Analysis, Aggregation Mechanism
\end{IEEEkeywords}

\section{INTRODUCTION}
Networks are extensively used in many real-world scenarios to represent various objects and their relationships, such as social networks\cite{b01,b02}, protein networks\cite{b03,b04}, knowledge graphs\cite{b05,b06} and citation networks\cite{b07}. Because it is difficult to carry out network analysis tasks directly on huge and complicated networks, network embedding is proposed to embed nodes into meaningful low-dimensional vectors in the euclidean space which capture intrinsic features of networks so that many downstream tasks, such as node classification\cite{b08,b09} and link prediction\cite{b10,b11,xu,xu2} can benefit from them (e.g., taking them as input to improve many downstream tasks).

\begin{figure*}[tbp]
\setlength{\belowcaptionskip}{-0.5cm}
\centering
\includegraphics[width=0.9\textwidth]{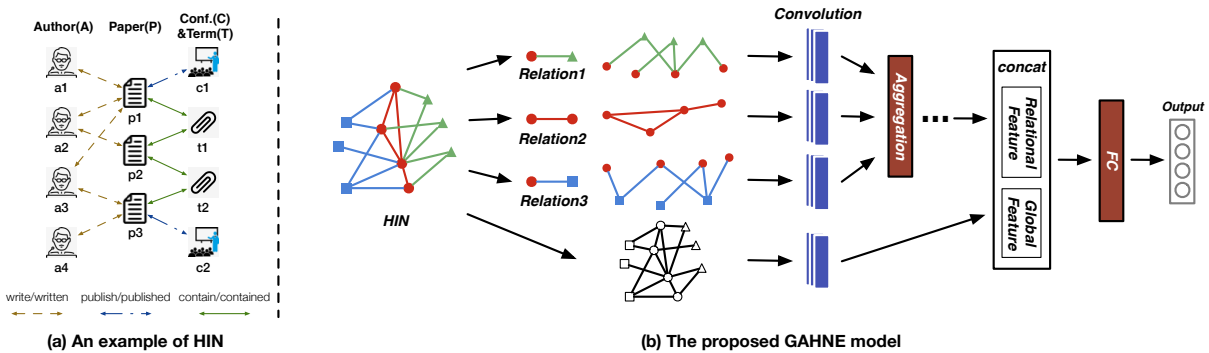}
\captionsetup{font={footnotesize}}
\caption{Overview of our work. (a) An illustrative example of a heterogeneous network (DBLP). (b) The overall architecture of GAHNE. Given an example of HIN which has three types of relations (color indicates the relation type). GAHNE firstly divides the HIN into single-type sub-networks and then aggregates the embeddings of nodes obtained in all sub-networks. Finally, GAHNE fuses the global information from the whole HIN (colourless network).}
\label{fig1}
\end{figure*}

In fact, real-world networks often compose of several types of nodes and edges (relations) with rich semantics, widely known as heterogeneous information network (HIN). Taking the extracted bibliography data from DBLP\footnote[1]{https://dblp.uni-trier.de/db/} in Fig.~\ref{fig1}(a) as an example, it consists of four types of nodes (i.e., paper, conference, term and author) and three types of edges (i.e., write/written relation, publish/published relation and contain/contained relation), which represent three different relations related to paper.

Opposite to the homogeneous network with a single type of nodes as well as a single type of edges, nodes in HIN have diverse type properties and different relation patterns among them. Obviously these additional information need to be considered when extracting features from HINs.
While there are already many network embedding approaches for homogeneous networks (e.g., factorization-based\cite{b12,b13}, random-walk-based\cite{b14,b15} and graph-neural-network-based\cite{b16,b08} methods), almost all these approaches have problems when facing HINs because they treat all nodes equally and ignore the influence of different relational semantics.

So far, considerable research work has implemented heterogeneous network embedding with the help of definitions of meta-paths, such as metapath2vec\cite{b17} and ESim\cite{b18}. On this basis, some others introduce graph neural network (GNN) to extract highly expressive features\cite{b19,b20,b21,b22}. Besides, of all GNN methods, graph convolutional networks (GCNs)\cite{b08} can effectively model structural dependencies. Some methods, such as GraphInception\cite{b23} and DHNE\cite{b24}, decompose HINs and adopt GCN to learn with obtained sub-networks separately. However, existing work more or less has the following shortcomings. Firstly, manual definitions of meta-paths in different HINs depend on specific domain knowledge and cause high costs. Secondly, models extract the features inside each type of semantic relations and then ignore the mutual effects between different types of nodes and relations, leading to suboptimal performance. Finally, node representations are just acquired from their related relational dependencies, causing to lose some aspects of information from global networks.

In order to address the above-mentioned issues, we propose a novel Graph-Aggregated Heterogeneous Network Embedding, named GAHNE. As Fig.~\ref{fig1}(b) shows, GAHNE employs node-level learning, channel aggregation and global feature fusion to generate node embeddings. First of all, according to the types of relations (edges), GAHNE divides the whole HIN into several single-type sub-networks, in which only one unique type of relations (edges) exists, which can remove the requirement of defining meta-paths. Since then, GAHNE obtains different semantic representations of nodes from internal channels by applying highly efficient and flexible graph convolution on each single-type sub-network.
After that, we introduce three aggregation mechanisms in GAHNE to adaptively integrate the representations from all channels in order to obtain the optimal combination of them. As a result, the mutual effects between different types of nodes and relations can be considered, which is helpful for completing feature representation comprehensively. Furthermore, for the sake of preserving the information from global networks, GAHNE fuses extra features obtained by graph convolution on the whole network to generate final node embeddings. Finally, the proposed model can be trained by back-propagation in an end-to-end manner. Our contributions are as follows:
\begin{enumerate}
\item We propose a novel Graph-Aggregated Heterogeneous Network Embedding (GAHNE) based on graph convolutional neural networks, which aims to extract the semantics of HINs as comprehensively as possible to improve the results of downstream tasks.
\item We elaborately design channel aggregation module with three candidate aggregators to aggregate semantic representations generated from different single-type sub-networks and global feature fusion module to fuse the global information into final embeddings.
\item We compare GAHNE with many state-of-the-art baselines on three real-world HIN datasets for semi-supervised node classification and unsupervised node clustering to show the effectiveness of node embeddings learned by GAHNE. Further analysis and visualization intuitively reveal the superiority of proposed model.
\end{enumerate}
\setlength{\parskip}{0.3em}
\section{THE PROPOSED MODEL}
\setlength{\parskip}{0em}
Before detailed introduction, we give some necessary definitions first. In a heterogeneous information network, denoted as $ \mathcal{G}=(\mathcal{V},\mathcal{E},\phi,\psi)$, where $\mathcal{V}$ indicates the object set and $\mathcal{E}$ indicates the link set. A heterogeneous graph is also associated with a node type mapping function $\phi:\mathcal{V}\to\mathcal{A}$ and a link type mapping function $\psi:\mathcal{E}\to\mathcal{R}$. $\mathcal{A}$ and $\mathcal{R}$ denote the sets of predefined object types and link types, where $\left|\mathcal{A}\right|+\left|\mathcal{R}\right|>2$. In this part, we describe our novel Graph-Aggregated Heterogeneous Network Embedding (GAHNE), as illustrated in Fig.~\ref{fig1}(b), which includes three parts: node-level learning, channel aggregation and global feature fusion.

\subsection{Node-Level Learning}
As mentioned in the previous section, HINs include several types of nodes and edges (relations). In order to obtain the feature embedding of a target node (i.e., node embedding), we should consider the influence of all kinds of relational semantics comprehensively. Hence, we process each of them independently at first. More specifically, we split the integral HIN $\mathcal{G}$ with $T$ types of relations into a series of sub-networks, denoted as $\{\mathcal{G}_t\mid t=1,2,\cdots,T\}$. As demonstrated in Fig.~\ref{fig1}(b), there are three single-type networks derived from the initial HIN. After this, we separately convolve and learn the deep semantic meanings of nodes in each obtained network via graph convolution operation. Kipf et al.\cite{b08} proposed the spectral graph convolution in the fourier domain basing on normalized graph laplacian $L^{sym}=I-D^{-\frac{1}{2}}AD^{-\frac{1}{2}}$, where $I$ denotes the identity matrix, $A$ denotes the adjacency matrix and $D = diag(\sum_{i}A(i,j))$. Corresponding to our problem, we can acquire the different relational semantics representations of a target node by multi-channel networks which respectively apply graph convolutions on these sub-networks.

To be specific, let $\{A_t\mid t=1,2,\cdots,T\}$ mean the adjacency matrices related to $\{\mathcal{G}_t\}$ with $N_t$ nodes $v^{i}_{t}\in\mathcal{V}_t$, edges $(v^{i}_{t},v^{j}_{t})\in\mathcal{E}_t$ and $D_{t} = diag(\sum_{i}A_t(i,j))$. Further, we use $P_{t}=D_{t}^{-1}A_t$, derived from random walk normalized laplacian instead of $L_{t}^{sym}$, since an asymmetric matrix is more suitable to define the fourier domain in view of the varied degree distribution in a HIN and the directed interaction between two connected nodes\cite{b24}. Moreover, compute the eigendecomposition: $P_{t} = U_{t} \Lambda_{t} U_{t}^{-1}$, where $\Lambda_{t}$ and $U_{t}$ denote the eigenvector matrix and the diagonal matrix of eigenvalues. In this way, spectral convolution on each single-relation network is defined as follows:
\begin{equation}
g_{\theta_{t}} \star X_{t} = U_{t}g_{\theta_{t}}U_{t}^{-1}X_{t} = U_{t}g_{\theta_{t}}(\Lambda_{t})U_{t}^{-1}X_{t}.
\end{equation}
The convolution is the multiplication of the input signal $X_{t} \in \mathbb{R}^{N_{t} \times D}$ of the network $\mathcal{G}_t$, where $D$ represents the input feature dimension, with a filter $g_{\theta_{t}} = diag(\theta_{t})$ parameterized by $\theta_{t} \in \mathbb{R}^{N_{t}}$, which can be understood as a function of the eigenvalues of $P_{t}$ (i.e., $g_{\theta_t}(\Lambda_t)$). $U^{-1}_{t}X_{t}$ is the graph fourier transform of input signal. Defferrard et al.\cite{b26} proposed that $K^{th}$-order polynomial Chebyshev polynomials can well-approximate $g_{\theta_{t}}$ as follows:

\begin{equation}
g_{\theta_{t}}(\Lambda_{t}) = \sum_{k=1}^{K}\theta_{t}^{k}\Lambda_{t}^{k},
\end{equation}
where $\theta_{t} \in \mathbb{R}^{K}$ is now a vector of Chebyshev coefficients. From Eq. 2 we have:
\begin{equation}\label{con:eq3}
g_{\theta_{t}} \star X_{t} = U_{t}(\sum_{k=1}^{K}\theta_{t}^{k}\Lambda_{t}^{k})U_{t}^{-1}X_{t} = \sum_{k=1}^{K}\theta_{t}^{k}P_{t}^{k}X_{t},
\end{equation}
\begin{equation}
H_{t} = \sigma(\sum_{k=1}^{K}P_{t}^{k}X_{t}\Theta_{t}),
\end{equation}
\setlength{\parskip}{-0.3em}
where $\Theta_{t} \in \mathbb{R}^{D\times d}$ denotes the trainable weight matrix which maps the input feature dimension $D$ to $d$ shared over the whole network $\mathcal{G}_t$ and $H_{t} \in \mathbb{R}^{N_{t}\times d}$ indicates the output matrix. We use an activation function $ReLU(\cdot)$ to add nonlinearity to output signals.
\subsection{Channel Aggregation}
In the previous section, we have learned the feature information of nodes on each sub-network. The final embeddings involve feature information in different relational scales that have unequal effects. Therefore, we need to aggregate output embeddings from all channels to obtain more comprehensive vector representations. Supposing $\{h_{t}^{v} \mid t = 1,2,\cdots,T\}$ is the set of output embeddings of a certain node $p_{v}\in \mathcal{V}$ generated from all $T$ channels in node-level learning. If a node does not belong to a sub-network, we use a zero vector to make complement. The aggregated embedding $z^{v} \in \mathbb{R}^{d}$ of $p_{v}$ can be defined as:
\setlength{\parskip}{0em}
\begin{equation}
z^{v} = Aggregation(\{h_{1}^{v}, h_{2}^{v}, h_{3}^{v}, \cdots, h_{T}^{v}\}),
\end{equation}
where $Aggregation(\cdot)$ denotes the special aggregator to carry out this purpose. Fig.~\ref{fig1}(b) cites a simple instance of this process. Here we introduce three choices of qualified candidate aggregators:
\begin{itemize}
\item \textbf{Attention-based Aggregator.} Inspired by some researches on attention mechanism in convolutional neural network (CNN)\cite{b33}, we propose a novel channel attention mechanism to measure the significance of feature information from every channel. This mechanism first makes non-linear transformation for the channel-wise embedding $h_{t}^{v}$ and then multiplies transformation result by a parameterized attention vector $q \in \mathbb{R}^{d_{q}}$ to get the importance of specific channel-wise embedding, where $d_{q}$ is the dimension of attention vector. Afterwards, we can calculate the sum of all the importance and then acquire the numerical value $w_{t}$ that represents the importance of sub-network $t$ in the final embedding. The formula of $w_{t}$ is as follows:

\begin{equation}
w_{t} = \sum_{v \in \mathcal{V}}q^{\mathsf{T}}\cdot Tanh(W\cdot h_{t}^{v} + b),
\end{equation}
where $W \in \mathbb{R}^{d_{q} \times d}$ and $b \in \mathbb{R}^{d_{q}}$ are non-linear transformation parameters. Next, the weight of each sub-network $\mu_{t}$ can be deduced by a softmax normalization as follows:
\begin{equation}
\mu_{t} = \frac{exp(w_{i})}{\sum_{i=1}^{T}exp(w_{i})}\label{eq5}.
\end{equation}
In particular, for $\forall t \in \{1, 2, \cdots, T\}$, we have $\mu_{t}^{1} = \mu_{t}^{2} = \cdots = \mu_{t}^{N}$, where $N$ is the number of nodes and $\mu_{t}^{v}$ denotes the weight of channel $t$ for node $v$.
In this way, we can aggregate these embeddings to the multi-channel embedding $z_{v}$ as follows:
\begin{equation}
z^{v} = \sum_{t=1}^{T}\mu^{v}_{t}h^{v}_{t}.
\end{equation}
\item \textbf{Gated Aggregator.} Gate unit\cite{b35} can effectively control the flow of previous information. Inspired by relevant thought, this mechanism proposes a gated module, which selectively aggregates rich scale features. The parametered gated vectors filter all dimensions of channel-wise embeddings:
\begin{equation}
g^{v}_{t} = Sigmoid(W'_{t}\cdot h_{t}^{v}),
\end{equation}
\begin{equation}
z^{v} = \sum_{t=1}^{T}g^{v}_{t} \odot h^{v}_{t}.
\end{equation}
Here $W'_{t} \in \mathbb{R}^{d \times d}$ is the parameterized training vector and $g^{v}_{t} \in \mathbb{R}^{d}$ indicates the gate obtained by the sigmoid function which controls the flow of information from $h^{v}_{t}$ and the scale of each component in $g^{v}_{t}$ is $[0,1]$.

\item \textbf{Pooling Aggregator.} The final aggregator adopts mean-pooling operator to aggregate all channel-wise embeddings. In this pooling approach, each channel’s output of a node is input in a fully-connected layer. Following this feature processing, we apply an element-wise mean-pooling operation to aggregate information across the channel set:
\begin{equation}
z^{v} = mean(\{ReLU(W^{pool}h_{t}^{v}+b^{pool}),\forall t \in \{1, 2, \cdots, T\}),
\end{equation}
where $mean$ denotes the element-wise mean operator and $ReLU(\cdot)$ is the activation function. This approach is inspired by advancements in GraphSAGE\cite{b16}. Our aggregator firstly computes features in the channel set and then uses the mean-pooling operator to effectively capture different aspects of the channel set.
\end{itemize}

Furthermore, let $Z_t$ indicates the aggregated embeddings of nodes in $\mathcal{G}_t$ and we provide a multi-layer convolutional neural network with the following layer-wise propagation rule:
\begin{equation}
H_t^{(l+1)} = \sigma(\sum_{k=1}^{K}P_tZ_t^{(l)}\Theta^{(l)}_{t}).
\end{equation}
Here, $\Theta^{(l)}_{t}\in \mathbb{R}^{d^{(l)} \times d^{(l+1)}}$ and $H_t^{(l+1)}\in \mathbb{R}^{N \times d^{(l+1)}}$ are specific trainable weight matrix and output signals, in which $d^{(l)}$ denotes feature dimension of $l^{th}$ layer. $Z^{(l)} \in \mathbb{R}^{N \times d^{(l)}}$ is the matrix of aggregated input embeddings in the $l^{th}$ layer; $Z_t^{(0)} = X_t$.
\setlength{\parskip}{-0.3em}
\subsection{Global Feature Fusion}
High-quality feature representation requires HINs to leverage both the homologous and heterogeneous information. Although we have performed the convolutions separately on each sub-network, we also lack a global scope of structural and semantic feature representation on the whole HIN.
To solve this challenge, we design an overall fusion layer shown in Fig.~\ref{fig1}(b). The input of this layer contains aggregated sub-network embeddings after $m$ layers $Z^{(m)} \in \mathbb{R}^{N \times d^{(m)}}$ and the embeddings $Z_{w}\in \mathbb{R}^{N \times d^{(m)}}$ learned on the whole graph. We can obtain $Z_{w}$ by using traditional GCN network as usual. In this layer, we explicitly fuse $Z^{(m)}$ and $Z_{w}$ to hold both partial and overall information:
\setlength{\parskip}{0em}
\begin{equation}
E_{f} = \left[Z^{(m)}, Z_{w}\right],
\end{equation}
here $E_{f} \in \mathbb{R}^{N \times 2d^{(m)}}$ is the concatenation of these two matrices, which carries more information. Following this, we use a fully-connected layer to provide more flexibility and non-linearity to learn the fused embeddings:
\begin{equation}
E = \sigma(W^{fc}E_{f}+b^{fc}).
\end{equation}
\setlength{\parskip}{-0.3em}
Now we have the final embeddings $E \in \mathbb{R}^{N \times d^{(m)}}$ of all nodes. And $W^{fc}$, $b^{fc}$ denote the weight matrix and bias of the layer.
\subsection{Model Training}
The final embeddings we obtain can be used as the input of the downstream tasks, and then design specialized loss function. We use a fully-connected layer and a softmax activation function $\sigma$:
\begin{equation}
F = softmax(E\Theta'),
\end{equation}
\setlength{\parskip}{0em}
here supposing the target node has $C$ class and $\Theta' \in \mathbb{R}^{d^{(m)} \times C}$ denotes the dimension reduction transformation matrix and $F \in \mathbb{R}^{N \times C}$ is the final probability matrix in which $F_{ij}$ indicates the probability that the $i^{th}$ node belongs to class $j$. Therefore, we can optimize the model weights by minimizing the cross entropy between the ground-truth and the prediction over a small fraction of labeled nodes iteratively:
\begin{equation}
Loss = -\sum_{v\in \mathcal{V}_{l}}\sum_{c=1}^{C} Y^{v}\lbrack c\rbrack\cdot\ln(F^{v}\lbrack c\rbrack),
\end{equation}
where $\mathcal{V}_{l}$ is the set of labeled nodes, $Y^{v}$ is the one-hot vector indicates the ground-truth labels of nodes and $F^{v}$ are the embeddings of labeled nodes. With the guide of labeled data, we can optimize the model via back propagation and learn the embeddings of nodes.

\section{EXPERIMENTS}
\subsection{Experimental Settings}

\linespread{1}
\begin{table}[t]
\captionsetup{font={footnotesize}}
\caption{Statistics of the datasets.}
\begin{center}
\begin{tabular}{cccc}
\toprule[0.9pt]
\textbf{Statistics}&\textbf{DBLP}& \textbf{Yelp}&\textbf{MovieLens}\\
\hline
\textbf{$\sharp$ Nodes} &$37791$&$3913$&$13119$ \\
\textbf{$\sharp$ Edges} & $170794$& $38668$ & $57932$ \\
\textbf{$\sharp$ Training} & $400$($9.8\%$) & $400$($15.3\%$)& $600$($16.3\%$) \\
\textbf{$\sharp$ Validation} &$200$($4.9\%$)& $200$($7.7\%$)& $300$($8.2\%$) \\
\textbf{$\sharp$ Test} &$3457$&$2164$& $2772$ \\
\toprule[0.9pt]
\end{tabular}
\end{center}
\label{table1}
\end{table}
\begin{enumerate}[leftmargin=*]
\item \textbf{Datasets.}
\begin{table*}[tb]
\captionsetup{font={footnotesize}}
\caption{Experiment results ($\%$) for the node classification task. (bold: best, underline: runner-up)}
\begin{center}

\resizebox{0.99\textwidth}{!}{
\begin{tabular}{|c|c|c||c|c|c|c|c|c|c||c|c|c|}
\hline
Datasets&Metrics & Train$\%$&Deepwalk&HIN2Vec&metapath2vec&GAT & HAN &GCN& DHNE & \tabincell{c}{GAHNE$_{atte}$} & \tabincell{c}{GAHNE$_{gate}$}  & \tabincell{c}{GAHNE$_{pool}$}  \\
\hline


\multirow{8}{*}{DBLP}&\multirow{4}{*}{Macro-F1}&$20\%$& 87.35 & 85.12 & 89.82 & 93.53 & 92.58 & 92.31 & 92.27 & \underline{94.17} & \textbf{94.32} & 93.23\\
& & $40\%$& 88.19 & 87.48 & 89.67 & 93.42 & 92.28 & 92.25 & 92.34 & \underline{94.10} & \textbf{94.46} & 93.31\\
& & $60\%$& 88.93 & 88.98 & 89.57 & 93.58 & 92.00 & 92.67 & 92.76 & \textbf{94.63} & \underline{94.52} & 93.54\\
& & $80\%$& 89.15 & 89.81 & 88.99 & 93.35 & 92.70 & 92.52 & 92.53 & \underline{94.28} & \textbf{94.48} & 93.57\\
\cline{2-13}
&\multirow{4}{*}{Micro-F1}& $20\%$& 87.91 & 85.68 & 90.41 & 94.02 & 92.14 & 92.82 & 92.95 & \underline{94.43} & \textbf{94.72} & 93.69\\
& & $40\%$& 88.71 & 87.98 & 90.26 & 93.88 & 92.78 & 92.74 & 92.99 & \underline{94.45} & \textbf{94.83} & 93.76\\
& & $60\%$& 89.51 & 89.51 & 90.25 & 94.04 & 92.54 & 93.16 & 93.42 & \underline{94.70} & \textbf{94.96} & 93.99\\
& & $80\%$& 89.72 & 90.29 & 89.72 & 93.88 & 93.20 & 92.86 & 93.00 & \underline{94.54} & \textbf{94.77} & 93.90\\
\hline

\multirow{8}{*}{Yelp}&\multirow{4}{*}{Macro-F1}&$20\%$& 62.28 & 69.25 & 67.62 & 68.68 & 69.58 & 71.96 & 70.93 & \underline{72.25} & 72.08 & \textbf{72.96}\\
& & $40\%$& 66.11 & 70.35 & 70.02 & 70.21 & 70.33 & 73.00 & 72.18 & \underline{74.16} & 73.24 & \textbf{74.89}\\
& & $60\%$& 67.61 & 71.33 & 71.55 & 71.02 & 69.83 & 72.57 & 72.45 & \textbf{74.18} & 73.05 & \underline{74.12}\\
& & $80\%$& 67.70 & 71.51 & 72.26 & 70.82 & 70.91 & 73.80 & 71.42 & \underline{74.61} & 73.40 & \textbf{75.99}\\
\cline{2-13}
&\multirow{4}{*}{Micro-F1}& $20\%$& 72.81 & 76.06 & 74.83 & 75.31 & 75.17 & 76.85 & 76.86 & 77.25 & \underline{77.33} & \textbf{77.65}\\
& & $40\%$& 74.99 & 76.68 & 76.00 & 76.25 & 75.17 & 77.35 & 77.63 & \underline{78.69} & 78.02 & \textbf{78.85}\\
& & $60\%$& 76.16 & 77.46 & 77.29 & 76.93 & 75.38 & 77.11 & 77.61 & \underline{78.70} & 77.58 & \textbf{78.38}\\
& & $80\%$& 75.62 & 77.71 & 77.40 & 76.67 & 75.93 & 77.94 & 77.15 & \underline{78.93} & 78.05 & \textbf{79.77}\\
\hline

\multirow{8}{*}{MovieLens}&\multirow{4}{*}{Macro-F1} & $20\%$& 56.53&53.41&52.50& 58.64 &52.24& 59.61&57.67& \underline{60.89} & 60.58 & \textbf{62.39}\\
&  & $40\%$& 59.94 & 56.90 & 55.52 & 60.42 & 52.26 & 60.64 & 58.58 & \underline{62.18} & 61.73 & \textbf{62.50}\\
&  & $60\%$& 60.79 & 59.26 & 58.33 & 60.05 & 52.42 & 61.38 & 57.57 & \underline{62.28} & 61.39 & \textbf{62.68}\\
&  & $80\%$& 62.55 & 61.63 & 60.30 & 61.01 & 53.69 & 61.74 & 60.14 & 62.79 & \underline{62.85} & \textbf{64.12}\\
\cline{2-13}
&\multirow{4}{*}{Micro-F1} & $20\%$ & 58.52 & 55.23 & 54.36 & 60.38 & 54.85 & 60.88 & 58.96 & \underline{62.27} & 61.99 & \textbf{63.45}\\
& & $40\%$& 61.63 & 58.46 & 57.55 & 61.92 & 55.12 & 61.86 & 59.87 & \textbf{63.57} & 63.08 & \underline{63.55}\\
& & $60\%$& 63.22 & 60.43 & 59.76 & 61.63 & 55.45 & 62.53 & 58.88 & \underline{63.32} & 62.64 & \textbf{63.66}\\
& & $80\%$& 64.04 & 62.79 & 61.77 & 62.50 & 55.70 & 63.03 & 61.46 & \underline{64.13} & 63.95 & \textbf{65.23}\\
\hline
\end{tabular}
}
\end{center}
\label{table2}
\vspace{-0.3cm}
\end{table*}
To evaluate the effectiveness of GAHNE, we use three real-world heterogeneous network datasets consisting of an academic network datasets DBLP, a social network Yelp\footnote[2]{https://www.yelp.com/dataset/} and a movie network MovieLens\footnote[3]{https://grouplens.org/datasets/movielens/}. The statistics of three datasets are summarized in TABLE~\ref{table1}. There are four types of nodes (paper (P), conference (C), author (A), term (T)), three types of edges (P-A, P-C, P-T) in DBLP and 4057 authors are divided into four research areas as labels. Yelp has five types of nodes (user (U), service (S), business (B), star level (L), reservation (R)), four types of edges (B-U, B-S, B-L, B-R), in which 2614 labeled businesses are from three categories. What's more, MovieLens is a film review dataset containing five types of nodes (movie (M), director (D), tag (T), writer (W), user (U)) with four types of relations (M-D, M-T, M-W, M-U) and there are three distinct genres in 3672 movies as labels.
\item \textbf{Baselines.}
We compare GAHNE against the following traditional, comparable, SOTA methods which aim to generate node embedding in supervised or unsupervised ways:
\begin{itemize}
\item[-] \textbf{DeepWalk\cite{b14}.} A random-walk-based network embedding method which is designed for the homogeneous networks without considering networks’ heterogeneity.

\item[-] \textbf{HIN2Vec\cite{b31}.} A heterogeneous information network embedding method via a deep neural network by considering the meta-paths.

\item[-] \textbf{metapath2vec\cite{b17}.} A metapath2vec model formalizes meta-path-based random walks and then leverages skip-gram model to perform node embeddings. We test on all meta-paths separately and report the one with the best results.

\item[-] \textbf{GAT\cite{b28}.} It is a semi-supervised neural network which considers the attention mechanism on the homogeneous graphs.

\item[-] \textbf{HAN\cite{b19}.} A heterogeneous graph neural network based on the hierarchical attention, including node-level and semantic-level attentions.

\item[-] \textbf{GCN\cite{b08}.} A scalable approach for semi-supervised learning on graph-structured data based on graph convolutional networks without considering network heterogeneity. We use random walk normalized Laplacian for better performance.

\item[-] \textbf{DHNE\cite{b24}.} A semi-supervised heterogeneous network embedding method which adopts convolutional layer on each decomposed homogeneous networks and bipartite networks and then concatenates the output vectors of each node from all networks without considering the weights of different networks.

\item[-] \textbf{GAHNE$_{atte}$.} Our proposed graph neural network with attention-based aggregator.
\item[-] \textbf{GAHNE$_{gate}$.} Our proposed graph neural network with gated aggregator.
\item[-] \textbf{GAHNE$_{pool}$.} Our proposed graph neural network with pooling aggregator.
\end{itemize}
\item \textbf{Implementation Details.} For the proposed GAHNE, we randomly initialize parameters and use Adam to optimize the model with a maximum of 200 epochs (adopting early stopping with a patience of 30). In the comparative experiments, both convolutional network branches for the whole network and the sub-networks in our framework are two layers, besides, we simply set convolution parameter $K=1$ in K-order approximation and set the dimension of the attention vector $q$ in attention-based aggregator to $128$. We leave their discussions in the following experiments. As for GNNs models, including GAT, HAN, GCN, DHNE and GAHNE, we set the learning rate to $0.01$, the dropout rate to $0.5$ and the $L2$ penalty weight decay to $0.0005$. For a fair comparison, we set the embedding dimension to $64$ for all the above algorithms. What's more, we randomly divide the labeled nodes into training set, validation set and the remaining as the testing set by certain ratios. For random-walk-based methods containing DeepWalk, metapath2vec, and HIN2Vec, we set window size to $5$, walk length to $100$, walks per node to $40$, the number of negative samples to $5$. For above methods requiring meta-paths, we define the commonly used schemes ``APA'', ``APTPA'' and ``APCPA''\cite{b19} on DBLP, ``UBSBU''\cite{b34} on Yelp, ``DMTMD'' and ``DMUMD''\cite{b24} on MovieLens to guide modeling. In practice, we implement GAHNE with Tensorflow1.13\footnote[4]{https://www.tensorflow.org/} to train model parameters and also use mini-batch gradient descent, which divides training data into several batches and updates parameters by each batch. In actual use, The label of the $i^{th}$ node can be predicted as $y^{i} = argmax(F^{i})$. Our code and data are available at website\footnote[5]{https://github.com/seanlxh/GAHNE}.
\end{enumerate}
\begin{table*}[t]
\captionsetup{font={footnotesize}}
\caption{Experiment results ($\%$) for the node clustering task. (bold: best, underline: runner-up)}
\begin{center}
\resizebox{0.95\textwidth}{!}{
\begin{tabular}{|c|c||c|c|c|c|c|c|c||c|c|c|}
\hline
Datasets&Metrics&Deepwalk&HIN2Vec&metapath2vec&GAT&HAN&GCN& DHNE & \tabincell{c}{GAHNE$_{atte}$} & \tabincell{c}{GAHNE$_{gate}$}  & \tabincell{c}{GAHNE$_{pool}$}  \\
\hline
\multirow{2}{*}{DBLP}&NMI& 78.69 & 73.06 & 75.38 & 80.59 & 72.95 & 76.88 & 76.93 & \underline{81.23} & \textbf{82.21} & 80.69\\
&ARI& 83.54 & 79.21 & 80.66 & 85.45 & 78.38 & 82.92 & 81.18 & \underline{86.15} & \textbf{87.12} & 86.41\\
\hline
\multirow{2}{*}{Yelp}&NMI& 34.08 & 35.70 & 36.03 & 33.34 & 38.93 & 39.66 & 39.08 & \underline{39.74} & \textbf{40.37} & 38.80\\
&ARI& 40.66 & 40.19 & 41.33 & 34.21 & 35.41 & 35.63 & \underline{43.08} & 42.76 & \textbf{43.77} & 41.11\\
\hline
\multirow{2}{*}{MovieLens}&NMI& 5.53 & 1.53 & 4.17 & 15.29 & 9.88 & 18.60 & 11.24 & \underline{21.14} & 17.34 &\textbf{21.82}\\
&ARI& 5.60 & 0.98 & 4.22 & 17.48 & 14.21 & 16.54 & 8.59 & \underline{20.66} & 19.42 & \textbf{22.13}\\
\hline
\end{tabular}
}
\end{center}
\label{table3}
\vspace{-0.3cm}
\end{table*}

\subsection{Node Classification}
Here we compare the effectiveness of different methods by the semi-supervised node classification task on testing set nodes via KNN classifier ($k$ = 5) with varying training proportions. In order to eliminate variance, we repeat the process for 10 times and report the averaged Micro-F1 and Macro-F1 in TABLE~\ref{table2}. As can be seen, GAHNE achieves the best and stable performance. Except GAHNE, the performances have ups and downs on the GNN-based baselines (GAT, HAN, GCN, DHNE). However, our GAHNE has around $1-3\%$ performance gain over the best baseline in general, which indicates that aggregation and fusion modules contain more comprehensive information. For random-walk-based graph embedding methods, Deepwalk has a fluctuated performance and HIN2Vec performs better than metapath2vec on Yelp and MovieLens since leveraging multiple meta-paths. But they all evidently lose to GAHNE that can utilize the heterogeneous node features via GNN architecture.
Compared to convolution methods GCN and DHNE, GAHNE has around $1-4\%$ improvement because of integrating the information from different relations appropriately and providing the global information. The performance of HAN is not entirely desired maybe because it discards all intermediate nodes along the meta-paths and just considers two end nodes. On semantically complex networks, meta-paths cannot be fully utilized in this way. It is interesting to find that some methods, such as Deepwalk, HIN2Vec, GAT and DHNE only have good performance in parts of datasets. Mainly because they rely on the importance of selected meta-paths or the specific structures in different networks. Thus they easily lead to uneven classification accuracy of the classifier, or poor generalization ability.

\subsection{Node Clustering}
Here we also use the unsupervised clustering task to evaluate the quality of obtained embeddings from various methods. We employ K-means to perform node clustering, and calculate normalized mutual information (NMI) and adjusted rand index (ARI) from clustering results to evaluate these methods. The number of clusters in K-means is the same as the number of nodes' classes. We use the same model configurations as in the node classification and repeat the process for 10 times. TABLE~\ref{table3} shows the average results. We can find that GAHNE is almost superior to other baselines on all datasets. Also, GNN-based algorithms usually achieve better performance than the traditional heterogeneous models because of the advantages of the GNN models. Traditional methods may lose efficacy, particularly in MovieLens, because of the dirty labels of movies. GNN-based baselines overall have better average performances, but they all fall behind in some cases since ignoring the heterogeneity or global information. On Yelp, almost all baselines have good results, DHNE in particular, because the semantics of Yelp is relatively simple, e.g., service (S) and reservation (R) only have two nodes and adjacent nodes have high similarity. The benefits of semantic aggregation are not significant.
\begin{figure*}[tb]
\centering
\subfigure[metapath2vec]{
\captionsetup{font={footnotesize}}
\begin{minipage}{4cm}
\centering
\includegraphics[scale=0.24]{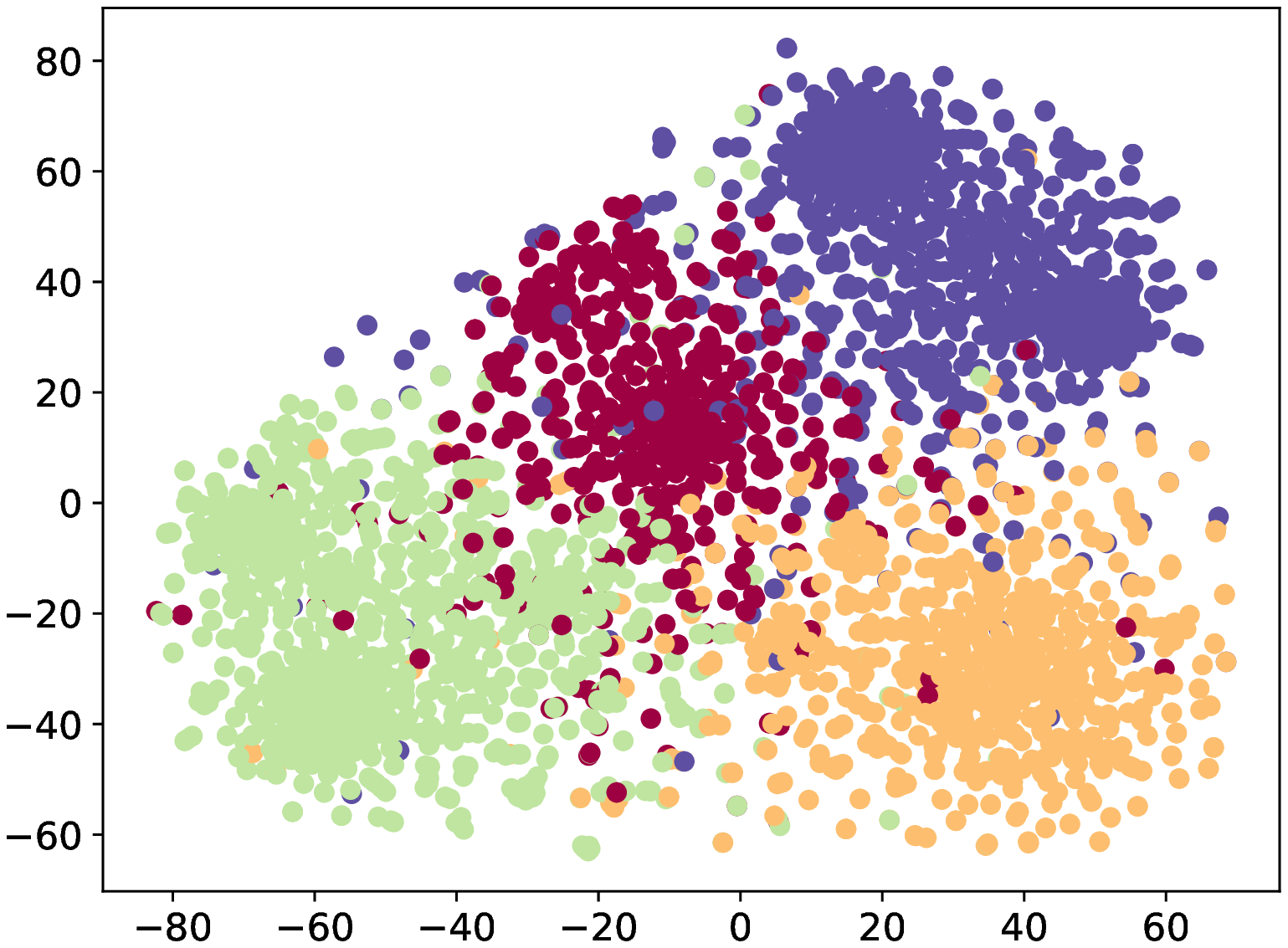}
\end{minipage}
}
\subfigure[GCN]{
\captionsetup{font={footnotesize}}
\begin{minipage}{4cm}
\centering
\includegraphics[scale=0.24]{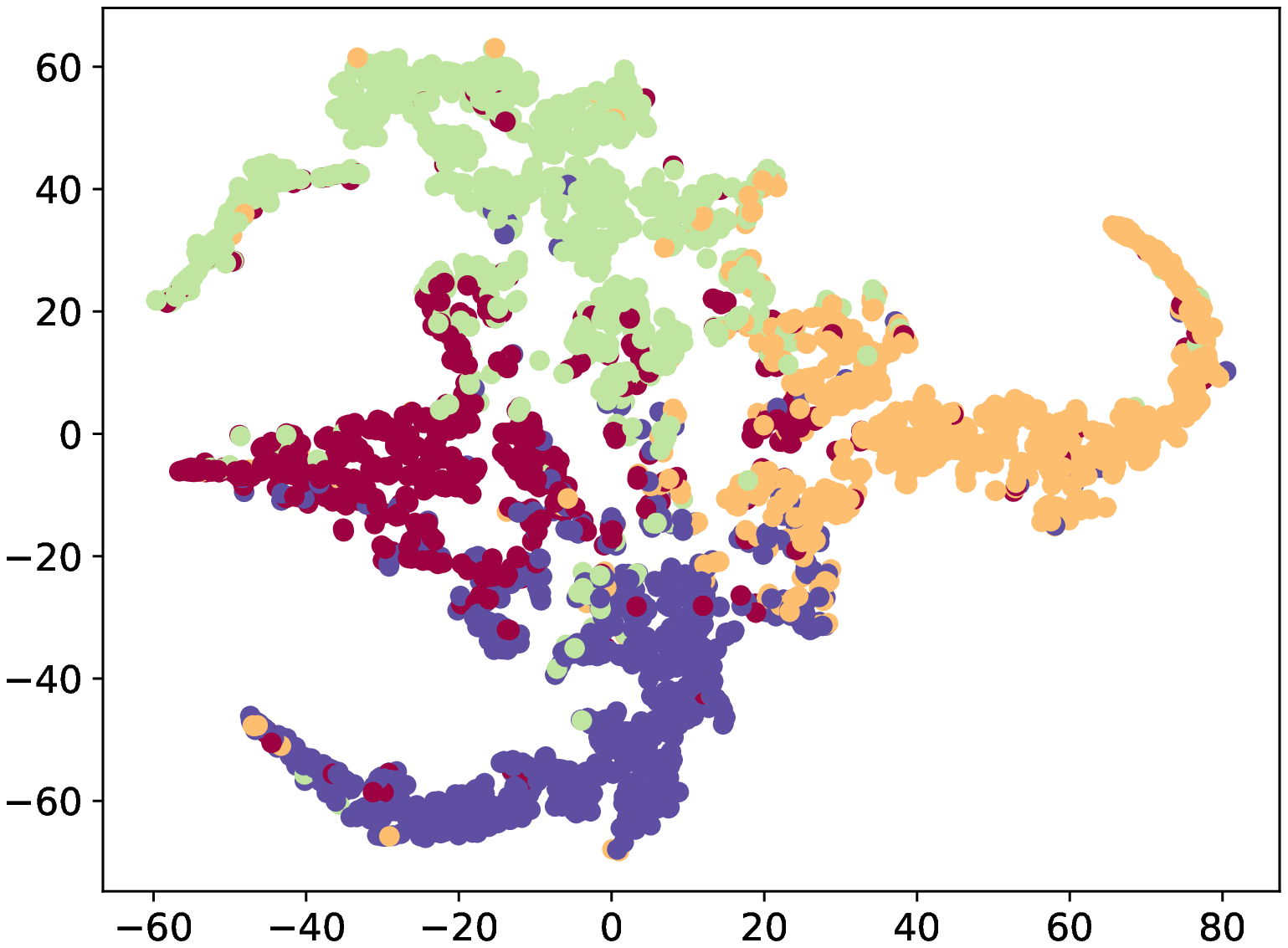}
\end{minipage}
}
\subfigure[HAN]{
\captionsetup{font={footnotesize}}
\begin{minipage}{4cm}
\centering
\includegraphics[scale=0.24]{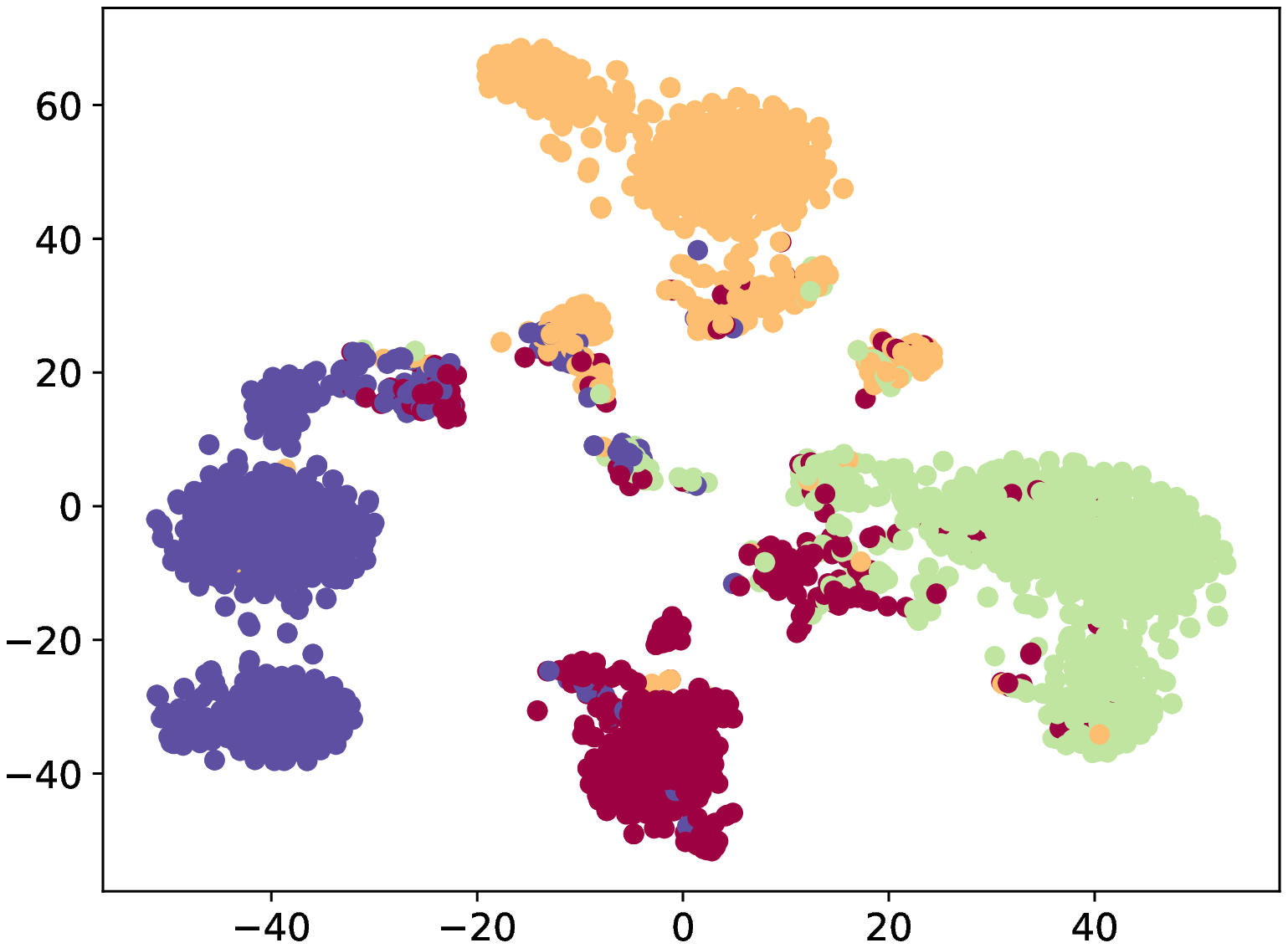}
\end{minipage}
}
\subfigure[GAHNE$_{atten}$]{
\captionsetup{font={footnotesize}}
\begin{minipage}{4cm}
\centering
\includegraphics[scale=0.24]{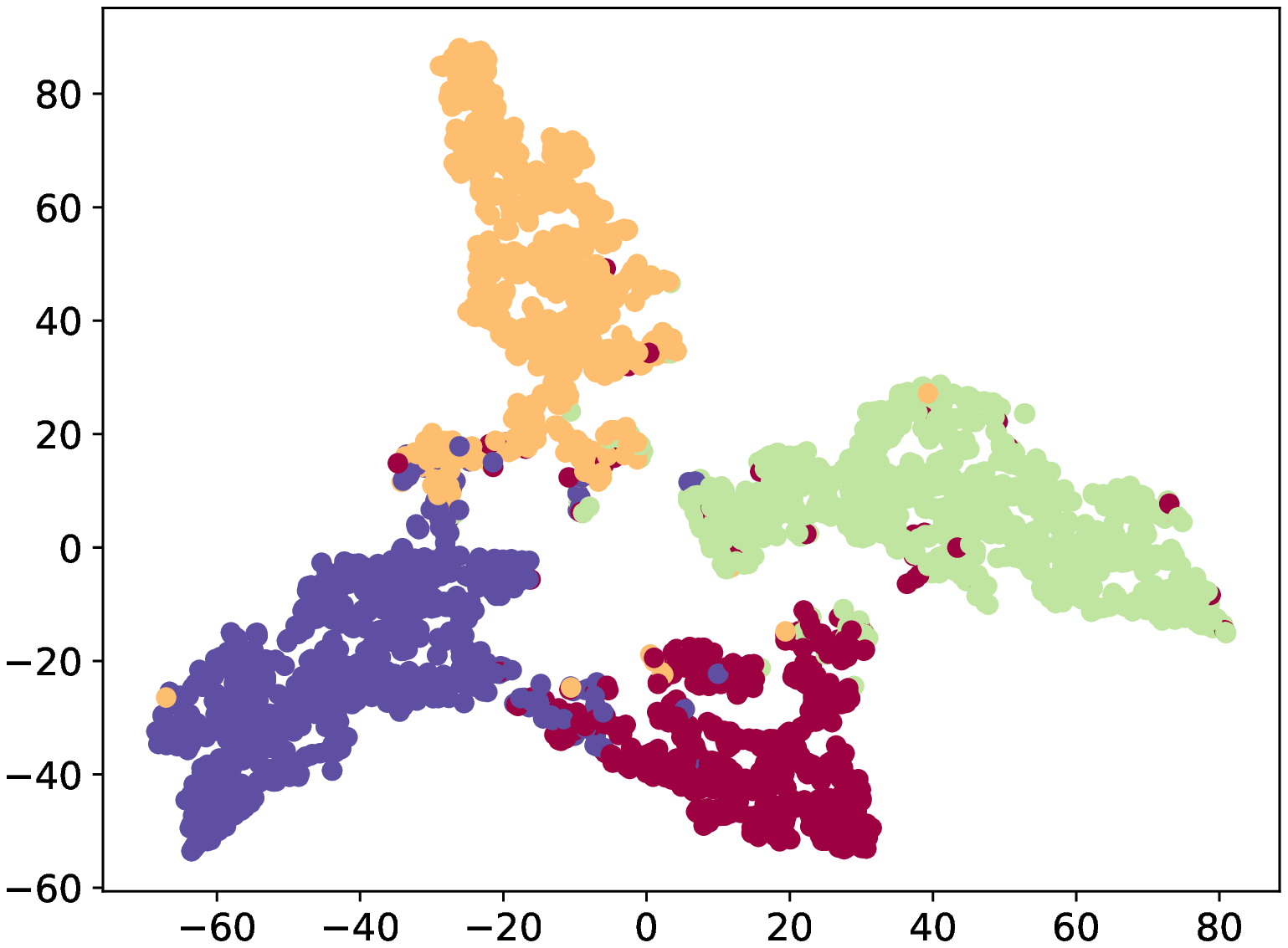}
\end{minipage}
}
\captionsetup{font={footnotesize}}
\caption{Visualization of authors' embeddings on DBLP. Different colors signify different research areas.}
\label{fig3}
\vspace{-0.5cm}
\end{figure*}

\begin{table*}[tb]
\captionsetup{font={footnotesize}}
\caption{Experiment results ($\%$) for ablation study (node classification task with $40\%$ training set and clustering task). The best performance is highlighted in boldface.}
\begin{center}
\resizebox{0.99\textwidth}{!}{
\begin{tabular}{|c||c|c|c|c|c|c|c|c|c|c|c|c|}
\hline
\multirow{2}{*}{Variant}&\multicolumn{4}{c|}{DBLP}& \multicolumn{4}{c|}{Yelp}& \multicolumn{4}{c|}{MovieLens} \\
\cline{2-13}
&Macro-F1&Micro-F1&NMI&ARI&Macro-F1&Micro-F1&NMI&ARI&Macro-F1&Micro-F1&NMI&ARI \\
\hline
GAHNE$_{atte}$& \textbf{94.30} & \textbf{94.53} & \textbf{81.23} & \textbf{86.15} & \textbf{73.80} & \textbf{78.39} & \textbf{39.74} & \textbf{42.76} & \textbf{62.04} & \textbf{63.32} & \textbf{21.12} & \textbf{20.66} \\
\hline
GAHNE$_{w/dag}$& 93.03 & 93.53 & 77.74 & 83.61 & 73.56 & 78.09 & 39.55 & 38.36 & 61.26 & 62.45 & 18.87 & 15.84 \\
\
GAHNE$_{w/dfu}$& 93.92 & 94.33 & 79.71 & 84.78 & 70.32 & 76.13 & 39.06 & 40.63 & 56.64 & 58.07 & 11.14 & 3.42 \\
GAHNE$_{tra}$& 93.12 & 93.56 & 79.34 & 85.27 & 73.32 & 77.74 & 37.72 & 39.32 & 61.36 & 62.51 & 19.37 & 18.32 \\
GAHNE$_{sig}$ & 92.94 & 93.44 & 77.52 & 82.97 & 73.47 & 78.04 & 38.39 & 40.11 & 61.57 & 62.07 & 18.12 & 18.82 \\
\hline
\end{tabular}
}
\end{center}
\label{table4}
\vspace{-0.5cm}
\end{table*}
\setlength{\parskip}{-0.2em}
\subsection{Further Investigations}
\setlength{\parskip}{0em}
\begin{enumerate}
\item \textbf{Visualization.}
To further illustrate the effectiveness of GAHNE, we conduct a group of visualization experiments to make comparisons intuitively. We project the learned embedding into a 2-dimensional space using t-SNE. Here we illustrate the authors' embeddings in DBLP obtained by metapath2vec, GCN, HAN and GAHNE$_{atte}$ in Fig.~\ref{fig3}, in which different colors are assigned to signify different research areas. From this visualization, we can see that the visual pattern of GAHNE turns out best. More concretely, it has high intra-class similarity and the boundaries between different research areas are clearly defined. As a meta-path-based network embedding model, metapath2vec relies on single meta-path and its ability is limited, so that different nodes crowd together. Moreover, since GCN does not take heterogeneity into consideration, the boundaries between different research areas are still blurry. As for HAN, the authors in the same class are scattered in the center of the picture, because HAN discards all intermediate nodes along the meta-path leading to deviation of the results.
\item \textbf{Parameter Sensitivity Analysis.} In this section, we do sensitivity analysis to some main parameters in GAHNE and Fig.~\ref{fig4} shows the NMI and ARI curves on DBLP dataset.
\begin{figure}[tb]
\centering
\subfigure[Dimension of final embedding]{\label{fig4:1}
\captionsetup{font={footnotesize}}
\begin{minipage}[t]{0.2\textwidth}
\centering
\includegraphics[width=1.05\textwidth]{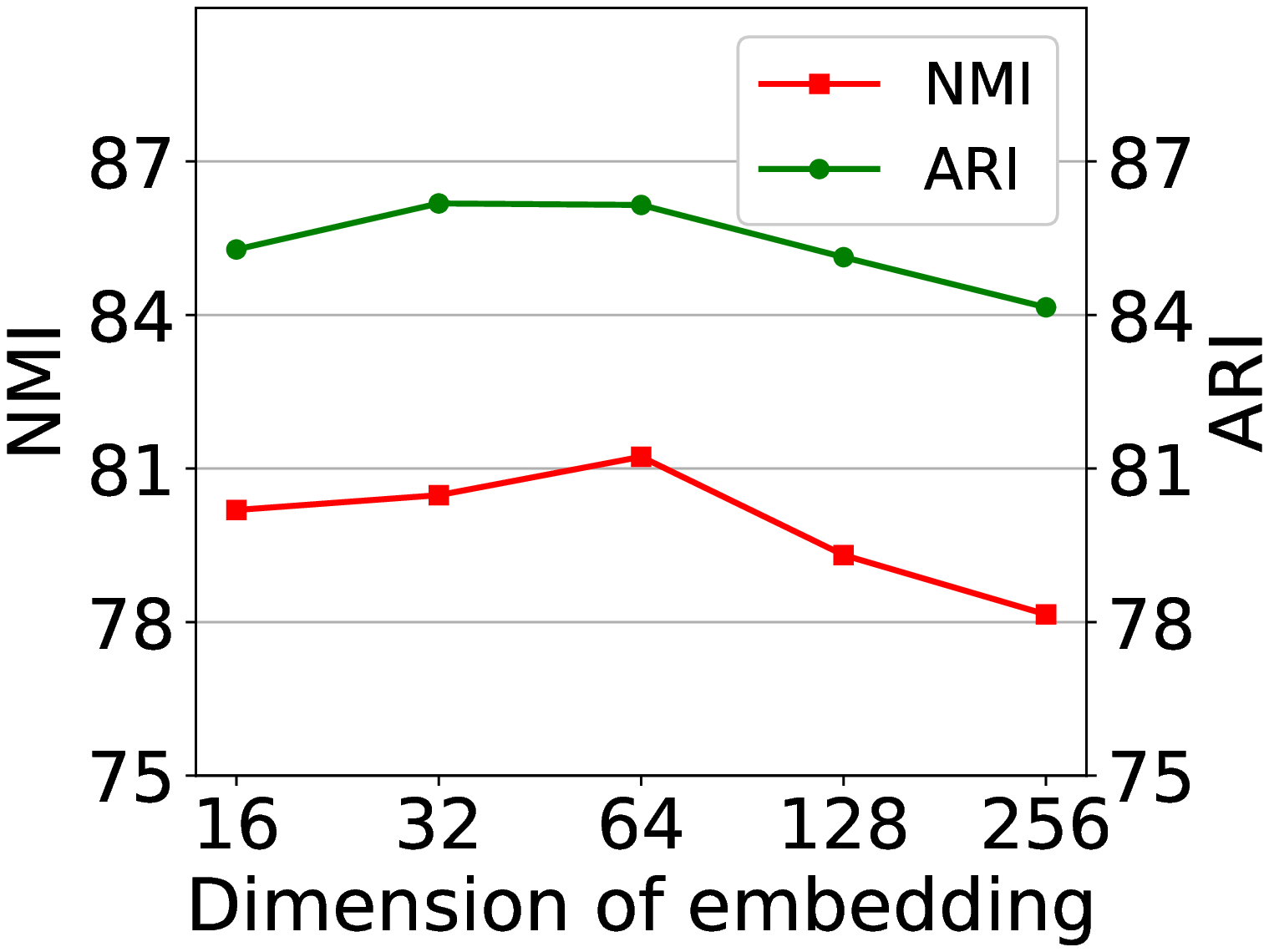}
\end{minipage}
}
\subfigure[Dimension of attention vector]{\label{fig4:2}
\captionsetup{font={footnotesize}}
\begin{minipage}[t]{0.2\textwidth}
\centering
\includegraphics[width=1.05\textwidth]{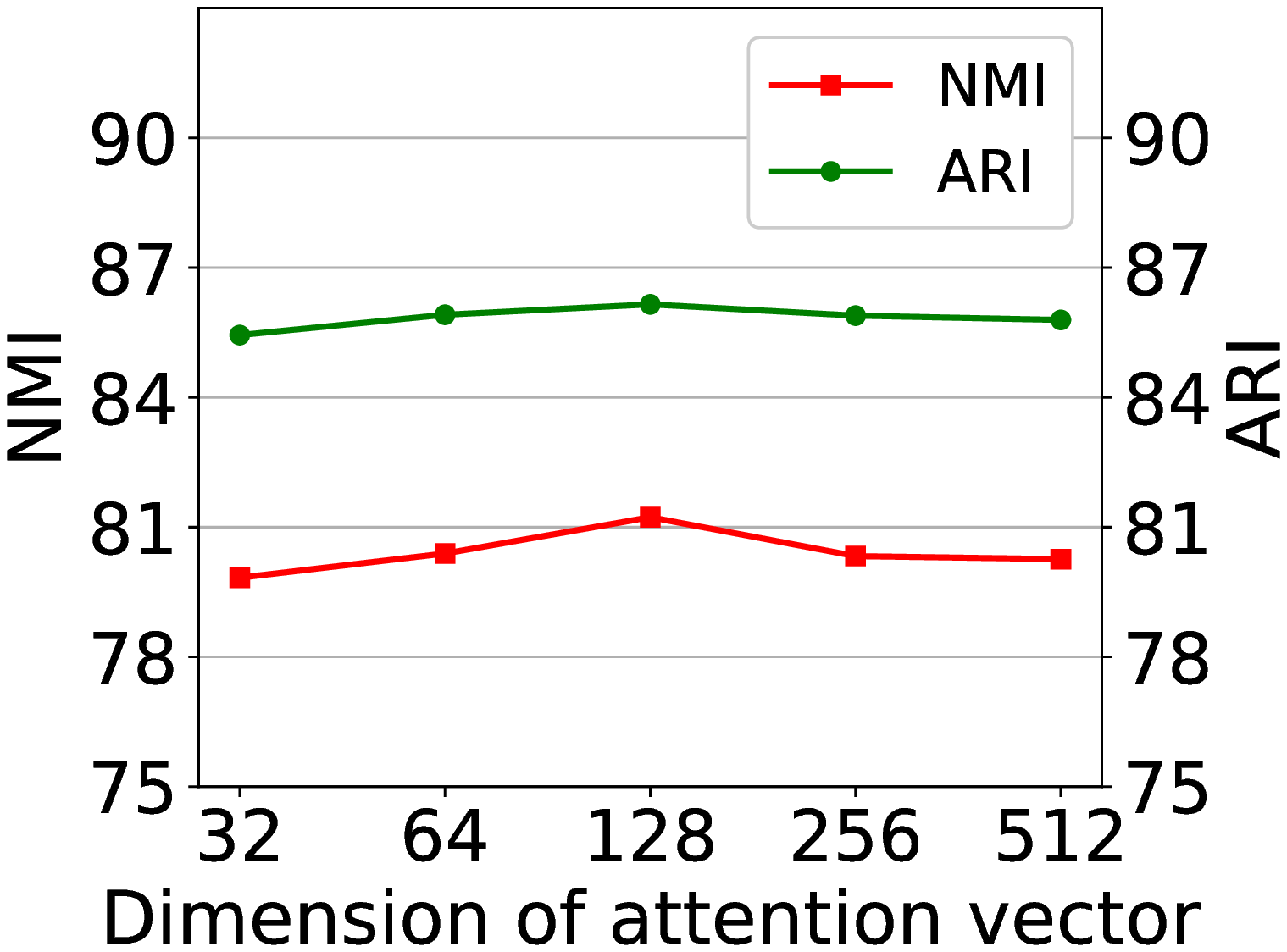}
\end{minipage}
}

\subfigure[Number of layers]{\label{fig4:3}
\captionsetup{font={footnotesize}}
\begin{minipage}[t]{0.2\textwidth}
\centering
\includegraphics[width=1.05\textwidth]{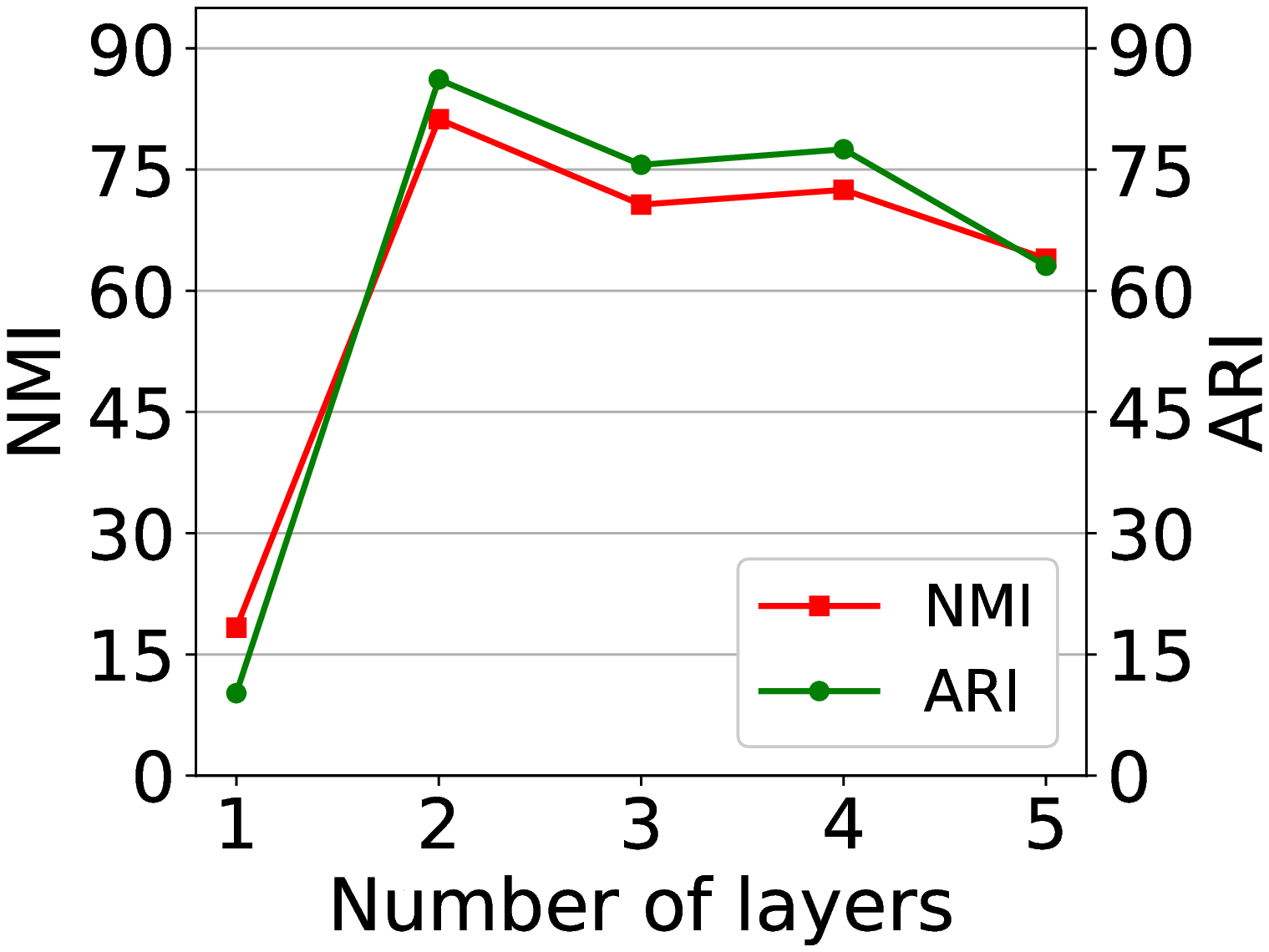}
\end{minipage}
}
\subfigure[Convolution parameter $K$]{\label{fig4:4}
\captionsetup{font={footnotesize}}
\begin{minipage}[t]{0.2\textwidth}
\centering
\includegraphics[width=1.05\textwidth]{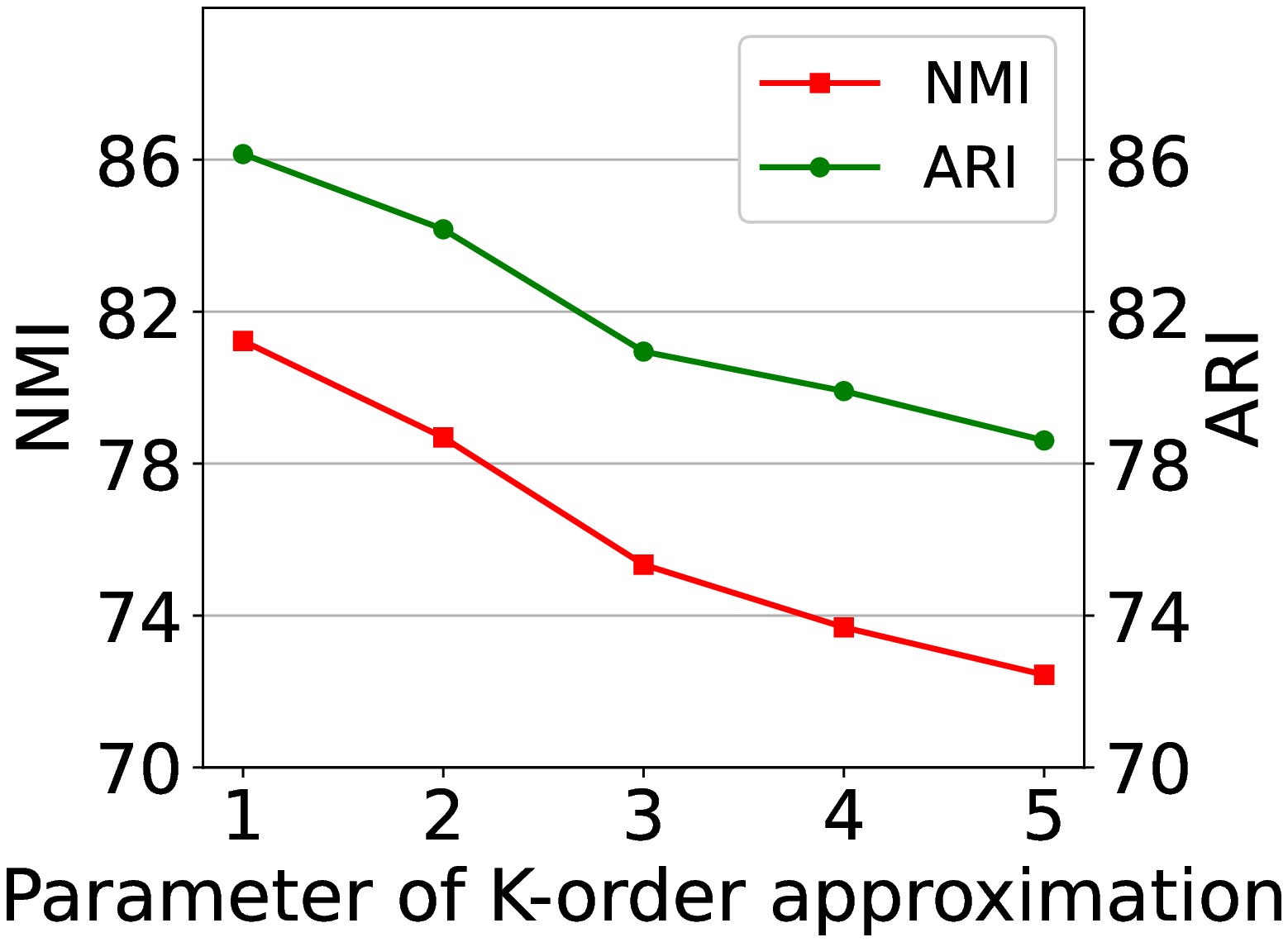}
\end{minipage}
}
\captionsetup{font={footnotesize}}
\caption{Parameter sensitivity of GAHNE.}
\label{fig4}
\end{figure}
\begin{enumerate}
\item \textbf{Dimension of the final embedding $F$.} We first test the effect of the dimension of the final embedding $F$, shown in Fig.~\ref{fig4:1}. As dimension grows, the performance raises steadily first and then shows a continued decline. The optimal performance is obtained when dimension is 64.

\item \textbf{Dimension of attention-based attention vector $q$.} Here we test how the dimension of the attention vector $q$ influences the effectiveness of attention-based aggregator. From Fig.~\ref{fig4:2}, we can conclude that our model is not sensitive to this parameter in general and the performance reaches the peak when the dimension is set to 128.

\item \textbf{Layer number of graph convolutional networks.} We compare the outcomes in different numbers of convolutional networks layers and report them in Fig.~\ref{fig4:3}. We can draw a conclusion that two-layered structure works best. A deeper network may bring bad effects because of over-smooth.

\item \textbf{Convolution parameter $K$.} We also investigate the effect of $K-order$ approximation of localized spectral filters on networks reported in Fig.~\ref{fig4:4}. Based on the results, we can find that limiting the layer-wise convolution operation to $K = 1$ works best. The information from more than one hop away from the target node offers no income, besides, we already have a two-layer structure.
\end{enumerate}
\end{enumerate}
\setlength{\parskip}{-0.5em}
\subsection{Ablation Study}
\setlength{\parskip}{0em}
In order to prove that all components of GAHNE are valid, we employ extra experiments on various variants of GAHNE. TABLE~\ref{table4} lists the results on the three datasets. Note that the proportion of training set is $40\%$ and the hyper-parameters in all convolutional networks are consistent. We denote GAHNE$_{w/dag}$ as the reference model which switches to calculate the average across the multi-channels relation embeddings instead of employing effective aggregation mechanisms. The decline in its performance, especially in clustering task, proves that our aggregation mechanism works. GAHNE$_{w/dfu}$ is an incomplete model that excludes global fusion. There is a significant decline in performance on MovieLens, which testifies the feasibility of global feature fusion. GAHNE$_{tra}$ is the contrast model by replacing multi-channels GCNs on different relations with a traditional GCN on whole networks, so that the whole framework has two same GCN branches. It confirms that capturing network features from different semantics helps improve GAHNE in a manner. GAHNE$_{sig}$ just selects a single relation embedding from multi-channels which produces the best results. The network information it captures is not comprehensive, so it causes a decline. All above ablation experiments prove the value of key parts of the proposed model.
\section{RELATED WORK}
\subsection{Graph Neural Networks}
Graph neural networks (GNNs) are classical models widely used in recent years, which transform the complicated input graph-structure data into meaningful representations for downstream mining tasks by information passing and aggregation according to dependencies in networks. Among all GNNs, graph convolutional networks (GCNs) are thought to become a dominating solution, falling into two categories: spectral\cite{b08,b25,b26} and non-spectral\cite{b27,b28} methods. As for spectral domains, Bruna et al.\cite{b25} proposed to utilize fourier basis to perform convolution in the spectral domain. ChebNet\cite{b26} introduced that smooth filters in spectral convolutions can be well-approximated by K-order Chebyshev polynomials. Kipf et al.\cite{b08} presented a convolutional architecture via a localized first-order approximation of spectral graph convolutions which further constrains and simplifies the parameters of ChebNet\cite{b26}. On the other hand, non-spectral methods are defined directly on the graph, operating on the target node and its spatial neighbors, so as to realize the convolution operation on the graph-structure. For example, Hamilton et al.\cite{b27} proposed GraphSAGE which generated embeddings by sampling and aggregating features from nodes' local neighborhood. In addition, there are many works utilizing attention layers in neural networks, such as GAT\cite{b28}, which leverages masked self-attention to enable specifying different weights to different nodes in a neighborhood. However, these methods do not deal with diverse types of nodes and edges specifically in the process of implementation, for which they are not suitable for HINs.
\subsection{Heterogeneous Network Embedding}
So far, heterogeneous network embedding learns low-dimensional vector representations of nodes mainly from the view of meta-paths semantic information. ESim\cite{b18} accepts user-defined meta-paths as guidance to learn vertex vectors in a user-preferred embedding space. Metapath2vec\cite{b30} produces meta-path-based random walks and then utilizes a heterogeneous skip-gram model to perform node embedding. HIN2Vec\cite{b31} is designed to capture the rich semantics contained in HINs by learning latent vectors of nodes and meta-paths. Some methods extend GNNs for modeling heterogeneous networks. HAN\cite{b19} proposes a graph neural network based on the hierarchical attention, including node-level and semantic-level attentions. MAGNN\cite{b32} makes improvements by enabling each target node to extract and combine information from the meta-path instances connecting the node. Besides, GraphInception\cite{b23} and DHNE\cite{b24} convert HINs into homogeneous networks in accordance with types of nodes and edges and then adopt graph convolution to learn with obtained networks separately. Following this, models concatenate all output vectors to get the final results. Nevertheless, all of the heterogeneous graph embedding methods introduced above have at least one of the following limitations: depending on manual meta-path selection process, ignoring the mutual effects between different types of nodes and relations, or discarding information from global networks.
\setlength{\parskip}{-0.3em}
\section{CONCLUSION}
In this paper, we mainly focus on processing the complex semantic information in heterogeneous information networks and give a heterogeneous graph neural network model based on graph convolution. The proposed model first converts a HIN into series of single-type sub-networks and then captures the semantic information from them respectively. After that, we provide some novel aggregation mechanisms to integrate them effectively. In addition, we fuse the supplement information from the whole network to mine networks comprehensively and stably. Extensive experiments prove that our final proposed approach GAHNE outperforms state-of-the-art methods in different areas with consistent level of performance. In the future, we will explore other possible measures to aggregate semantics and fuse the global information in order to better capture the structural information of HINs.\\[3pt]
\textbf{Acknowledgement.} The work was supported by the National Key Research and Development Program of China (No. 2019YFB1704003), the National Nature Science Foundation of China (No. 71690231), Tsinghua BNRist. Lijie Wen is the corresponding author.

\end{document}